\documentstyle[12pt]{article}
\begin{document}

\begin{tabbing}
\hskip 11.5 cm \= {CERN-TH-7389/94}\\
\hskip 1 cm \>{August 1994}\\
\end{tabbing}
\thispagestyle{empty}
\vskip 0.3cm
\begin{center}
{\Large\bf WHITHER DO THE MICROLENSING}
\vskip 0.3cm
{\Large\bf BROWN-DWARFS}
\vskip 0.3cm
{\Large\bf ROVE?}
\vskip 1.0 cm
{\bf A. De R\'ujula, G. F. Giudice, S. Mollerach and E. Roulet}
\vskip .3cm
CERN,\\
1211 Geneva 23\\
Switzerland

\vskip 1. cm
{\large ABSTRACT}
\end{center}

\noindent{The EROS and MACHO collaborations have reported  
observations of light curves of stars in the Large Magellanic Cloud  
that are compatible with gravitational microlensing by intervening  
massive objects, presumably Brown-Dwarf stars.  The OGLE and MACHO  
teams have also
seen similar events in the direction of the galactic Bulge.
Current data are insufficient to decide whether the Brown-Dwarfs are  
dark-matter constituents of the non-luminous galactic Halo, or belong  
to a more conventional population, such as that of faint stars in the  
galactic Spheroid, in its Thin or Thick Disks, or in their possible  
LMC counterparts. We discuss in detail how further observations of  
microlensing rates, and of the moments of the distribution of event  
durations, can help resolve the issue of the Brown-Dwarf location,  
and eventually provide information on the mass function of the dark  
objects.}

\vskip 0.5 cm

\noindent{CERN-TH.6272/91}

\noindent{August 1994}
\newpage

\section{Introduction}

In a paper uneagerly published in 1936, Einstein (1936) described  
how the apparent luminosity of a star may be temporarily amplified
 by  
the gravitational field of a second star that crosses close to the  
line of sight to the first. He concluded that the effect was of no  
practical interest. Half a century later, Paczy\'nski (1986)  
revised Einstein's conclusion by noting that the search for  
microlensing events in the direction of the {\it Large Magellanic  
Cloud} (LMC) may resolve the question of the extent to which the mass  
of the dark Halo of our galaxy is due to {\it Brown-Dwarfs,} stars  
too faint to be readily observed otherwise.

The EROS (Aubourg {\it et al.} 1993) and MACHO 
(Alcock {\it et al.} 1993) collaborations have reported  
``microlensing'' events of the expected characteristics in the  
direction of the LMC, and the MACHO (Alcock {\it et al.} 1994) and 
OGLE 
(Udalski {\it et al.} 1994) groups  
have seen similar events in the direction of the {\it ``Galactic  
Bulge'',} centrally located in the Milky Way. In spite of the  
currently low statistics, the observers have established that the  
temporal shape, achromaticity, location in the Hertzsprung-Russell  
diagram, and non-repetitive character of the events place them far  
away from the tails of the distributions inferred from the observed 
variable stars. The events are either due to a novel and most  
implausible type of variable, whose properties are ``just so''  
as to fool the observer, or to microlensing by the faint interlopers 
that one is searching for.

We are primarily interested in Brown-Dwarfs, {\it compact objects}  
that act as gravitational lenses but that may otherwise be directly  
observable, with considerable toil, only at infrared wavelengths  
(Kerins \& Carr 1994).
Not to shine at visible frequencies, Brown-Dwarfs must be {\bf  
lighter} than the thermo-nuclear ignition threshold, which for a  
compact object predominantly made of H  
and He is $m_{max}\sim 0.085 M_\odot$ (Graboske \& Grossman 1971). 
Not to have evaporated during the age of the  
Galaxy, these bodies must be {\bf heavier} than $(10^{-6}$-$10^{-7})
M_\odot$ (De R\'ujula, Jetzer and Mass\'o 1992).

There is a cutoff on the mass of Brown-Dwarfs more stringent  
than the evaporation limit: the {\it ``Jeans''} mass $m_{min}\sim  
7\times 10^{-3} M_\odot$ (Low \& Lynden-Bell 1976), the lowest  
mass of a detached H/He gas-cloud fragment whose self-gravitational  
collapse into a compact object is not prevented by the developing  
thermal pressure. Although the Jeans limit is not absolute (Jupiter,  
Saturn, Uranus and Neptune are counterexamples to it), the relatively  
long durations of the observed microlensing events indicate that the  
objects responsible for them are in the upper domain of allowed  
masses, the question of a precise lowest-mass cutoff appears to be 
rather moot. We shall consequently restrict our considerations to 
compact  
objects whose masses range from the Jeans mass to the  
hydrogen-burning threshold. On-going observations are sensitive to  
microlensing by objects with masses lying anywhere in this  
mass range.

The existence of a dark halo in spiral galaxies is decisively  
demonstrated by their observed rotation curves. The nature  
of the constituents of this dark mass, contrary-wise, is debatable  
and debated. An important role in the discussion is played by  
constraints on the universal average ``baryonic'' density, stemming  
from the relative abundances of primordial elements that are  
inferred, after significant corrections, from spectral observations  
(Walker {\it et al.} 1991). 
Recent measurements of the $^2$H abundance in distant  
intergalactic clouds (Songaila {\it et al.} 1994, 
Carswell {\it et al.} 1994) 
favour the conclusion that most  
baryonic matter is visible and accounted for, leaving no room for a  
predominantly  (baryonic) Brown-Dwarf constituency of our Halo. 

Often used as an argument against a Halo made of Brown-Dwarfs 
is the contention that the density of all other stellar distributions 
falls off with distance from the galactic centres significantly 
faster than $\sim r^{-2}$, the behaviour required to explain a 
flat rotation curve. But Sackett {\it et al.} (1994) have recently 
observed that the Halo of the spiral NGC5907, though very dim, has 
a luminosity distribution that traces its dark mass.

In our opinion, it is too early to accept as conclusive the  
arguments favouring a baryonic or non-baryonic composition of  
galactic halos. We discuss the direct search for Brown-Dwarfs in the 
Halo of the Galaxy, if only because failure to find them in  
significant amounts would constitute the most intriguing  
observational result: it would strongly advocate for a Halo  
consisting of a substance more subtle than the ones we are made of.

The interest of Brown-Dwarfs would transcend stellar physics, should  
they significantly contribute to the Halo dark mass. Thus the spirit  
in which we shall discuss the microlensing data: Brown-Dwarfs in the  
Galaxy's Halo will be considered the ``signal'', dim lensing bodies  
from other stellar populations will be regarded as ``backgrounds''.

Various locations for these backgrounds have been discussed: the {\it  
Thin Disk}, the {\it Thick Disk} (Gould, Miralda-Escud\'e \& Bahcall 
1994) and the {\it  
Spheroid} (Giudice, Mollerach \& Roulet 1994), all of them stellar 
populations extending beyond the location of our own solar system. 
Though we are immersed in these stellar distributions,  
the contribution of their unseen astral bodies to microlensing in
 the direction of the  
Magellanic Clouds was once thought to be quite negligible,  
relative to the full signal of a Brown-Dwarf-dominated Halo. But, 
if the dark constituency of these galactic components is sizable, 
they may significantly contribute to the LMC and SMC lensing rates. 
They could even account for the observations (Gould,  
Miralda-Escud\'e \& Bahcall 1994, Giudice, Mollerach \& Roulet  
1994) that, within very poor statistics, seem to fall short of the 
expected Halo rates. Dim objects in the LMC itself may also contribute 
to the microlensing of LMC stars. 

To visualize the invisible, we show in  Fig. 1 contour plots of the 
different galactic dark-mass distributions that we shall discuss. 
The $x$ and $y$ axis are in the galactic plane,  the $y$ axis digs into  
the figure. For each population, the inner and outer contours
correspond to volume densities of $10^{-2} M_\odot/\mbox{pc}^3$ and
$10^{-3} M_\odot/\mbox{pc}^3$. The solar position is at ($-8.5$,\,
0,\, 0) kpc. The densest and most convenient microlensing
target-fields of source stars are the Magellanic Clouds and the
galactic Bulge. The LMC, assesing its distance from us to be
$55$ kpc, is at (0,\, $-$46,\, $-$29) kpc. The SMC, reckoned to be
located at $65$ kpc, is at (17,$-$39,$-$45) kpc. The galactic
Bulge fills the central 1--2 kpc. 
We devote Section 2 to a detailed discussion of the spatial distributions
of visible and dark mass in our Galaxy and in the LMC. 

In this paper we discuss simple strategies to analyse microlensing  
data of modest statistics. We combine the modelling of various signal  
and background stellar populations with the information reflected in  
the first few moments (De R\'ujula, Jetzer \& Mass\'o 1991) 
of the event-duration  
distribution: the  
number of events (the ``zeroth'' moment); the average duration  
$\langle T \rangle$; and the time dispersion 
$\Delta T/T\equiv \sqrt{\langle T^2\rangle -\langle  
T\rangle^2}/\langle T\rangle$.
The mass function of the lenses (number of objects per unit mass  
interval) is unknown apriori, and the predictions for the various  
duration moments depend on its assumed functional form. We deal with  
this problem by ``sweeping'' our results over the very large domain of  
mass-function ansatze discussed in Section~3. 

In Section 4 we recall the basic expressions for the microlensing rate 
as a function of event duration, in the divers viewing directions of 
interest, and for the various lensing populations. In Section 5 we 
develop a feeling for the results by discussing ideal observations 
of single-mass lensing objects. In Section 6 we discuss realistic 
limited-statistics observations of objects whose mass function and 
location are unknown. 
We illustrate the effects of the detection efficiency as a function of
 event duration. This is not always possible, since only OGLE has 
already published the relevant  
information and MACHO is currently analysing the problem in 
detail\footnote{ We are indebted to Pierre Bareyre for discussions 
on the EROS efficiencies.}.

The microlensing predictions corresponding to the Halo signal 
and the various possible backgrounds can be conveniently 
visualized by plotting the event rate and the event time-dispersion
 against the mean event duration. For observations in the direction 
of the  
Magellanic Clouds, even after the ``blurring'' induced by the apriori 
ignorance of the lensing-objects' mass function, the signal and 
backgrounds occupy  
distinct regions of these plots. It is therefore quite conceivable 
that, as the statistics improves to a  
few dozen events, the data favour one of 
these regions and the culprit  
microlensing star population can be pinned down. This is our main  
result, which will be 
reflected in Figs.~4 and 5. The comparison of SMC and LMC 
observations would also help locating the lensing agents.

In the direction of the Bulge the microlensing observables predicted 
for the different dark
populations overlap to some extent, once the ignorance of the lensing
mass function is duly taken into account. Thus, our proposed procedure 
to select the stellar distribution to which the microlensing objects 
belong should work for the LMC, but not for the Bulge. There the 
situation is even more challenging for, as is almost always the 
case in astronomy, the observations have spoiled the simplicity 
of the original picture. 

The Bulge event rates  
reported by the OGLE and MACHO 
collaborations are larger than expected in any simple Brown-Dwarf scenario. 
And the event durations are surprisingly long.
The observed rate can be brought to agree with theory by combining the 
contributions of several stellar populations. But the rate and the event 
durations cannot be simultaneously described by any combination of the 
conventional Halo, Disk and Spheroid dark-matter models (plus the 
contribution of faint Disk stars). To quantify the severity of this 
problem, and given the scarcity of the data, we must resort to a 
Kolmogorov--Smirnov test of the time distribution.
We conclude that the disagreements are serious enough to justify 
modifications of ere accepted models of the inner galactic realm. 
Our analysis and some of these modifications 
(Alcock {\it et al.} 1994, Paczy\'nski {\it et al.} 1994) are 
specified in Section 6d. 
 
The titles and subtitles of Sections 2--6 image the organization 
of the paper. In the conclusions of Section 7 we fail to be truly 
conclusive, for our current understanding of galaxies is still 
modest, and the microlensing observations are still infants.

\section{Galactic components}

In this Section we summarize the hypothesis and empirical constraints  
concerning the dark mass of the Galaxy and of the Magellanic Clouds.

$a$) {\it The galactic Halo}

We assume the density-profile of our galactic Halo to be
spherical\footnote{ A flattening of the Halo would be of little relevance
to LMC observations, but would increase the optical depth towards the
SMC by up to 40\% (Sackett \& Gould, 1993). The effect of tilting the
axis of a non-spherical Halo has been discussed by Frieman \&
Scoccimarro (1994).} and to correspond to a rotation curve flattening
at $v_H\simeq 220$ km/s.
Let $R_\otimes \simeq 8.5$ kpc be our distance to the galactic centre
and $a\sim 3$ kpc the Halo's ``core radius''.  We adopt a
halo density distribution: 
\begin{equation}
\rho\, (r)=\rho_{_H}\; {R_\otimes ^2+a^2\over r^2+a^2}\; .
\end{equation}
The uncertainty in $a$, which is rather large (Caldwell \&
Ostriker 1981 and Bahcall, Schmidt \& Soneira 1983), 
significantly affects
the microlensing predictions in the direction of the Bulge, but is
ineffectual in the direction of the Magellanic Clouds. The rotation
curve of our galaxy is only measured out to $r\sim 20$ kpc (Fich and
Tremaine, 1991), and microlensing in the LMC and SMC significantly
depends on the assumed Halo density at larger values of $r$: it is not 
possible to assign an empirical error to the predictions for Halo 
microlensing rates. 

The local halo density implied by fits to the Galaxy's rotation  
curve is $\rho_{_H}\sim 10^{-2} \, M_\odot/\mbox{pc}^3$. If and when a signal  
for microlensing by objects in our Halo is established, the main  
result of the observational campaigns would be the comparison of the  
dynamically-determined value of the dark-mass density with that extracted  
from the microlensing data. The ratio of this two independent  
measurements is the fractional contribution of compact objects to the  
Halo's dark mass. 

The velocities of the constituents of a gravitationally  
self-consistent spherical Halo are isotropic at any  
point, and follow a Maxwellian distribution (Chandrasekhar 1942):
\begin{equation}
f(v)={ 1 \over (2\pi)^{3/2} \; \sigma^3 } 
\; e^{-v^2/2\sigma^2}\; ,
\end{equation}
with a dispersion $\sigma=v_H/\sqrt{2}$.

The galactic Halo is overwhelmingly ``dark''.

\vskip .3cm

$b$) {\it The Milky Way's Populations
of Visible Stars}

   Most of the visible stars in our galaxy belong to a {\bf Thin  
Disk}. The disk's
mass density is exponentially distributed with a characteristic  
radial scale $r_{_D}\simeq 3.5$ kpc and a scale height of $H\simeq  
100$ pc for the very
young stars and gas, and $\simeq 325$ pc for older stars. The Thin  
Disk rotates at a constant $v_{tDisk}\simeq 220$ km/s and its velocity  
dispersion is small, of the order of 20 km/s near the Sun and  
somewhat larger towards the galactic centre.

The faint stars of the Thin Disk may  contribute to microlensing. As  
a typifying example, consider a main-sequence star of mass $\leq 0.3  
\, M_\odot$ located at a distance of 1 kpc: its apparent luminosity  
would be smaller than that of the stars being monitored in the LMC.  
Since the Thin Disk mass function (Scalo 1986) is known,  we shall  
explicitly and separately compute their microlensing  
effects\footnote{ We neglect White Dwarfs, whose contribution to the  
local column density is less than 20\% of that of main-sequence stars  
(Bahcall 1984).}.

Gilmore and Reid (1983) fitted observations of stellar photometry to
a superposition of two exponential disks, and found evidence for the  
existence of a new galactic component, the {\bf Thick Disk}, to which  
they attributed a scale height $H=1.5$ kpc and a dispersion velocity  
$\sigma_z \simeq 60$ km/s. It is now believed that these measurements  
were contaminated by stars in the Spheroid (to be discussed anon),  
and that a proper
treatment of the Thick Disk leads to $H\simeq 1$--1.2 kpc and  
$\sigma_z \simeq 40$ km/s (see for example Freeman 1987 and  
references therein). The Thick Disk rotates  more slowly than the  
Thin Disk, with $v_{\, TDisk}\equiv v_{tDisk}-v_{drift}$, where  
$v_{drift}\simeq 50$ km/s is the ``drift'' velocity relative to the  
Thin Disk's rotation.

The stars in the Thin Disk are of Population I: young and metal-rich.
The Thick Disk stars have intermediate metallicities and hence are
older than Thin Disk stars. The nature of the Thick Disk is
controversial, it may be a tail of the Thin Disk, formed as the older
stars were scattered to larger speeds; it may be a star population
born in some other way.

The existence of the  
{\bf Spheroid}, yet another stellar component of the Galaxy, 
is implied by star counts at high latitudes and by  
observations of nearby, high-velocity stars. Its density profile is  
approximately spherical, falling roughly as $r^{-3.5}$. The Spheroid  
is dominantly supported  by the large velocity dispersion of its  
constituents ($\sigma\simeq 120$ km/s) and has a negligible bulk  
rotation. It is made of old and metal-poor stars and believed to have  
been formed in the  first $10^9$ years of the Milky Way's  
``protogalactic'' life.

\vskip .3cm
$c$) {\it Dark Mass in the Disks and Spheroid}

To estimate the backgrounds to a putative microlensing signal by  
compact objects in the galactic Halo, we need to know the dark  
constituency of the observed stellar populations. This is something  
that, almost by definition, we do not know. There are, however, two  
handles on this problem. One concerns the extrapolation of the mass  
functions of observed stars downwards in mass into the region of  
faint or unlit objects. No such extrapolation is to be trusted; we  
only discuss this subject for its subliminal worth, letting the dark  
mass functions we employ in the microlensing predictions vary over a  
large domain of possibilities.

The second handle on the dark mass of the established stellar  
distributions concerns the comparison of their visible and  
``dynamical'' masses, and is more secure. The observed galactic  
rotation curve and velocity dispersions imply upper limits on the  
local column density $\Sigma$, the total mass in a cylinder of unit  
area extending perpendicularly to the galactic plane to an explored  
height of some $\pm 1.1$ kpc. Subtraction of the column density of  
observed stars translates into a normalization of the allowed local  
dark-mass density of a star population with an assumed spacial distribution. 
The use of this  
constraint is simplified by the fact that we are not after a precise  
number, but after a sort of ``maxi-estimate'' of the non-Halo  
dark-mass backgrounds to microlensing.
We shall normalize the dark mass of the Thin and Thick Disks,  
independently and in turn, to our maxi-estimate of the dark local  
column density. In the case of the Spheroid, which contributes little  
to $\Sigma$, we exploit directly the comparison of observations of  
its luminous and dynamical mass.

The mass function of the main sequence stars in the Thin Disk is  
known almost down to the hydrogen burning limit $M_{HB}\simeq 0.085  
\, M_\odot$ (Scalo 1986), and follows the ``Salpeter law''  
$dn/dm\propto m^{-2.3}$ for masses larger than $\sim 2 \, M_\odot$.  
Towards lower masses the mass function flattens, and a smooth  
continuation into the Brown-Dwarf domain ($m < M_{HB}$) might suggest  
a relatively small dark component of the Thin 
Disk\footnote{ A Thin Disk population of infrared-emitting 
Brown-Dwarfs is significantly bounded by the IRAS survey 
measurements (Kerins \& Carr 1994).}. 
Little is known about the mass function of Thick Disk stars.

Measurements of the Galaxy's rotation curve imply a local  
column density in a disk component, $\Sigma$, inferior to $\sim 100 \,  
M_\odot/\mbox{pc}^2$ (Binney \& Tremaine 1987). 
Kuijken and Gilmore (1989 a,b), from  
measurements of dispersion velocities at altitudes $|z|\!\! <\!\!  
z_{max}\! =\! 1.1 \, \mbox{kpc}$, infer a total (disk plus halo)  
column density $\Sigma\,\{|z|\!\! < \!\! z_{max}\}= (71\pm 6) 
M_\odot/\mbox{pc}^2$. The rotation-curve constraint implies that for  
the disk alone
$\Sigma_{\, Disk}\,\{|z|\!\! < \!\! z_{max}\}=(48\pm 9) M_\odot/  
\mbox{pc}^2$. Gould (1990) reanalyses the same data to obtain  
$\Sigma_{\, Disk}\,\{|z|\!\! < \!\! z_{max}\}=(54\pm 8)M_\odot/  
\mbox{pc}^2$. We choose as a conservative upper bound $\Sigma_{\,
Disk}\,\{|z|\!\! < \!\! z_{max}\} = 75 M_\odot/\mbox{pc}^2$
(Sackett and Gould (1993) make this choice for the total disk column
density, while Gould, Miralda-Escud\'e and Bahcall (1994) overpass the
quoted analysis to adopt $\Sigma_{\, Disk} = 100 
M_\odot/\mbox{pc}^2$). 

To obtain the maximum allowed dark-matter component, one must  
subtract from the total column density the various visible contributions. 
These are approximately $10 \, M_\odot/\mbox{pc}^2$ in gas, $30
\, M_\odot/\mbox{pc}^2$ in main sequence stars\footnote{ This is also  
what results
from the integration of  the Scalo mass function, corrected in order  
to account for binary stars as in Bahcall (1984).}, and $5\,  
M_\odot/\mbox{pc}^2$ in white dwarf stellar remnants and red giants 
(Bahcall 1984).  
For a Thin Disk essentially all the mass is within $|z|\! <\! 1.1$  
kpc, while for a Thick Disk of scale height $H\sim 1$ kpc only $\sim  
2/3$ of the column density would be within those heights.  
Accordingly, we choose for the total dark column densities  
$\Sigma_{\,Thin}=30M_\odot/\mbox{pc}^2$,
$\Sigma_{\,Thick}=45M_\odot/\mbox{pc}^2$.

Let $R=\sqrt{x^2+y^2}$ and $z$ be cylindrical coordinates for the  
radial and altitude coordinates of the Disks. We parametrize their  
dark-mass distributions as: 
\begin{equation}
\rho\, (R,z)=\rho_{_D} \; e^{\, (R_\otimes -R)/ r_{_D} } \;
e^{-|z|/ H} \; ,
\end{equation}
with $R_\otimes =8.5$ kpc, $r_{_D}\simeq 3.5$ kpc, and  
$\rho_{_D}=\Sigma_{dark}/(2H)$ the local dark-mass density. For the  
Thin and Thick Disk's scale heights we adopt $H=325$ pc and $H=1.2$  
kpc, respectively.

The luminous mass of the Spheroid is (1 to $3) \times 10^9 
M_\odot$ (Bahcall, Schmidt and Soneira 1983), while its dynamical  
mass is modelled to be (5 to $7)\times 10^{10} M_\odot$ (Caldwell  
and Ostriker 1981, Rohlfs and Kreitschman 1988). This indicates that  
a large fraction of the Spheroid is in non-luminous Brown-Dwarfs or  
stellar remnants. The mass function of Spheroid field stars increases  
steeply\footnote{ The luminosity function used by Richer and Fahlman  
(1992), and hence the results of their analysis, have been criticized  
by Bahcall {\it et al.} (1994).} towards low masses (Richer and  
Fahlman, 1992), behaving as $dn/dm\propto m^{-4.5\pm 1.2}$ in the  
interval extending from $0.14 \, M_\odot$, the minimum observed  
masses, to $0.5 M_\odot$. It suffices to extrapolate these  
observations down to $\sim 0.04M_\odot$ (half the hydrogen-burning  
limit) to get the Spheroid's total dynamical mass. In this sense  
Brown-Dwarfs are well motivated constituents of the dark mass of the  
Spheroid.

For the Spheroid's mass distribution we use a simplified expression,  
in good
agreement with the ones used in dynamical fits (Ostriker \& Caldwell 
1982, Rohlfs \& Kreitschman 1988):
\begin{equation}
\rho\, (r) = \rho_{_S}\, \left( { {R_\otimes^{1/2} +b^{1/2}} \over  
{r^{1/2}+b^{1/2}} }
\right)^7 \; ,
\label{Sphe}
\end{equation}
with $b=0.17$ kpc and $\rho_{_S}=1.5\times 10^{-3}
M_\odot/\mbox{pc}^3$ chosen to give, upon integration, a total dark  
mass of 5.7 $\times 10^{10} M_\odot$ in the Spheroid.
The inner 1 kpc of this Spheroid model, which mass is $\sim 
10^{10}M_\odot$, may be identified with the galactic Bulge (Blanco 
\& Tendrup 1989, Blitz \& Spergel 1991a)\footnote{ Some observations 
suggest an additional ``nuclear Bulge'' (Rich 1990, Kent 1992), 
presumably a galactic bar (Blitz and Spergel 1991b, 
Binney {\it et al.} 1991). Kiraga \& Paczy\'nski (1994) 
discuss this stellar distribution; in Kent's (1992) model 
it entails a microlensing rate 
comparable to that of the inner Spheroid considered by us.}.

\vskip .3cm

{\it d) Visible stars and dark-matter in the Large Magellanic Cloud}

May dim or lightless objects located in the LMC itself contribute to  
microlensing of the LMC stars? (Gould 1993, Wu 1994, Sahu 1994).
 In discussing this subject, as in  
the case of our Galaxy, it is useful to recall what is known about  
visible stars in the LMC.

The luminous component of the LMC is usually described as a flat  
exponential {\bf Disk} that is seen nearly face-on (determinations of  
the inclination angle range from $27^\circ$ to $45^\circ$).
The disk's radial scale is $r_{_{LMC}}\simeq 1.6$ kpc. The question  
of its scale height is more involved.

The LMC gas and star-clusters have a small velocity dispersion  
($\sigma<10$ km/s). ``Intermediate'' (more evolved) populations  
--$\!$ planetary nebulae, CH stars and intermediate long-period  
variables $\!$-- have $\sigma\simeq 20$ km/s, characteristic of a  
Disk population with a scale height $h\simeq 0.5$ kpc. The older  
population of long-period variables (OLPV) has a velocity dispersion  
of $\sigma\simeq 33$ km/s (Hughes, Wood and Reid 1991), corresponding  
more closely to a spheroidal distribution than to a disk-like one.  
Due to a partially-rotational support, this {\bf Spheroid} should be  
quite flattened, with a minor- to major-axial ratio $c/a=0.3$ to  
0.5. The spatial distribution resulting from the OLPV counts behaves  
as $\sim r^{-1.8}$.

The LMC's rotation curve is that of a solid body up to a radius of  
$\sim 2$ kpc. At larger distances the rotational velocity flattens to  
an approximately constant 80 km/s (Westerlund 1991 and references  
therein). The total mass of the LMC inside a radius of 6 kpc,  
inferred from the rotation curve, is $M_{LMC}\simeq 6\times 10^9
M_\odot$. Larger total masses, as high as $2\times 10^{10}
M_\odot$, are obtained if the flat rotation curve is extrapolated up  
to a distance of 15 kpc, as the velocity of three outlying globular  
clusters would suggest doing (Storm and Carney 1991, Schommer {\it et al.}  
1992). If this large mass is accounted for by a conventional  
spherical {\bf Halo}, its density distribution should behave as  
$r^{-2}$  and its dispersion would be $\sigma\simeq 80/\sqrt{2}$  
km/s.

We consider the possible additional contribution to the LMC  
microlensing rates of three different dark stellar populations in  
the LMC itself:

A spherical {\bf Halo} with a scale of $a=1$ kpc and a density profile:
\begin{equation}
\rho\, (r)=\rho_{_H}\;{a^2\over
r^2+a^2}\;\; \theta\, (r-r_{max})
\end{equation}
with $r_{max}=10$ kpc and a total dark-mass of $10^{10}M_\odot$,
chosen as compromises of the various bits of information. The
corresponding density is $\rho_{_H}=0.09 M_\odot/\mbox{pc}^3$.

A {\bf Disk} distribution similar to that of the ``intermediate''  
population objects. We choose its density profile, in cylindrical  
coordinates, to be:
\begin{equation}
\rho\, (R,z)= \rho_{_D}
\; e^{-R/r_{_D}} \;\; \mbox{sech}^2
\left({z\over h}\right) \, ,
\end{equation}
where $r_{_D}=1.6$ kpc, $h=0.43$ kpc and $\rho_{_D}=0.29 
M_\odot/\mbox{pc}^3$,
corresponding to an assumed total dark mass of $4\times 10^9 
M_\odot$ in the disk.

A flattened {\bf Spheroidal} distribution similar to that of the  
OLPV, normalized to a total dark mass of $4\times 10^9 M_\odot$  
inside $r_{_S}=6$ kpc. Its density function is assumed to be:  
\begin{equation}
\rho\, (R,z)=\rho_{_S}\;{b^2\over
b^2+\xi^2}\;\;\theta\, (\xi-r_{_S}) ,\ \ \ \  
\xi\equiv\sqrt{R^2+{z^2\over
\cos^2\Psi}},
\end{equation}
where the ellipticity is cos$\Psi\equiv c/a=0.5$, $b=1$ kpc, and  
$\rho_{_S}=0.09 M_\odot/\mbox{pc}^3$ for the presumed total dark  
mass.

None of the above LMC dark-mass scenarios
is defensible to a level of confidence comparable to the analogous  
scenarios in the Galaxy. We discuss them only to illustrate in some  
detail how their effects on microlensing are relatively small. When  
considering a possible addition of invisible LMC objects to those of  
the dark populations of the Galaxy, a certain degree of consistency  
is called for. It would be peculiar, for instance, to assume that the  
halo of the LMC is made of some inscrutable form of ``particle'' dark  
matter while that of the Galaxy is baryonic.

\section{Dark-Mass Mass Functions}

Let $dn (x) / dm$  be the dark-mass function: the number of somber objects  
with mass between $m$ and $m + dm$ at a given point $x$ in the Galaxy  
or in its Halo. Very little is known about the explicit $m$ and $x$  
dependence of this object, besides the constraints on its integral  
over all masses, discussed in the previous section. To explore  
microlensing predictions a measure of definiteness --$\!$ and of 
guesswork $\!$-- are required.

For every dark-mass population we make a {\bf factorization  
hypothesis}: $dn (x) / dm$ is the product of a distribution $dn_0/dm$  
depending only on $m$, times the pertinent dark-mass density profile:
\begin{equation}
\frac{dn(x)}{dm} = \frac{\rho\, (x)}{\rho_0} \;\frac{dn_0}{dm}\; ,
\label{FH}  
\end{equation}
where we have used a local density $\rho_0$ as a normalization. For
the very extensive galactic Halo this is a fairly strong conjecture,
while for the dark component of the various known stellar
distributions, that are multifarious, more localized and better
understood, the supposition sounds like an eminently reasonable
approximation.

The factorization hypothesis reduces our task to that of speculating  
on the functional form of $dn_0/dm$. We posit 
two families of  
ansatze, one very simple, the other quite sensible. 
The first hypothesis is that of mass functions peaked at a particular  
mass
$m_0$:
\begin{equation}
{d{n_0}\over dm} = \rho_o \;
\frac{\delta (m-m_0)}{m_0} \, ;
\label{delta}
\end{equation}
the second is that of a power-law distribution, of index $\alpha$, in  
the range $m_1<m<m_2$:
\begin{equation}
{dn_0\over dm} =\rho_o \; \frac{m^{-\alpha}}{D}
\; \theta\, (m-m_1)\;\theta\, (m_2-m)\, ,
\;\;\;\;\;\;
D\equiv {\int_{m_1}^{m_2} m^{1-\alpha}\, dm}\, , \label{power}
\end{equation}
with $\alpha$ any positive or negative number. We let $m_{i}$,  
$i=0,1,2$, lie anywhere in a range $m_{min}$ to $m_{max}$. A  
plausible value for $m_{min}$ is given by the Jeans limit, and we  
follow Low \& Lynden-Bell (1976) in choosing
$m_{min}=7\times 10^{-3}M_\odot$, though values as low as $4\times  
10^{-3}M_\odot$ are possible (Palla, Salpeter \& Stahler 1983).  
For $m_{max}$ we take a generous hydrogen-burning limit of
$m_{max}=10^{-1}M_\odot$.

In the study of microlensing observables, we let the dark-mass  
functions $dn_0/dm$ vary over the entire range of possible values of  
$m_i$ and $\alpha$.

\section{Microlensing Rates and Time Moments}

Let $L$ be the distance to a source that is microlensed by an object  
of mass $M$
crossing close to the line of sight at distance $x$ from the  
observer. Follow Einstein in defining a radius that characterizes the  
gravitational lensing scale:  
\begin{equation}
R_{_E}\equiv \sqrt{{G M (L-x)\, x \over
L\,c^2}}\, .
\end{equation}
Define $u$ as the distance of the lens to the line of sight in  
$R_{_E}$ units:
\begin{equation}
u^2(t)\equiv \left({l\over R_{_E}}\right)^2+\left({v^{\, \perp}    (t
-t_o)\over R_{_E}}\right)^2,
\label{time} 
\end{equation}
with $t_0$ the time of closest approach, $l$ the impact parameter,  
and $v^{\, \perp}$ the modulus of $\vec{v}^{\; \perp}$, the  
projection on the plane orthogonal to the line of sight of the  
relative velocity  between the lens and this very line (hereafter,  
the label ``$\perp$'' always refers to this projection).

The microlensing amplification of the luminosity of the source is:  
\begin{equation}
A\, [t]={u^2(t)+2\over u(t)\, {\sqrt {u^2(t) + 4}}}\; .
\end{equation}
The maximum amplification, $A_{max}=A\, (u_{min})$, occurs at time  
$t=t_0$, as $u$ reaches its minimum value $u_{min}\equiv u(t_0)=  
l/R_{_E} $.

The amplification curve $A\, [t]$ is universal, its shape is fully  
determined by its height $A_{max}$ and its duration $T\equiv R_{_E}  
/v^{\, \perp}$. For observations in a given direction, the  
distribution of events in $A_{max}$ tests the hypothesis that the  
distribution of impact parameters ought to be uniform (all the scale  
lengths characterizing the dark-mass distributions are much bigger  
than the conceivable Einstein radii). The distribution of event  
durations $T$ is pregnant with interesting but well convoluted  
information, which we aim to extract. The values of $t_0$ in 
Eq. (\ref{time}) are only  
relevant to the anthropological question of discovery priorities.

Let $u_{th} $ be the smallest value of $u_{min}$ for which the  
amplification is sufficient for an event to be detectable
(e.g. $u_{th}  =1$ corresponds to a detection amplification threshold  
$A_{max}>  1.34$). Define a microlensing tube of radius $u_{th}  \,  
R_{_E}$ around the line of sight. Consider a circle on the tube's  
surface and let $\alpha$ be the angle defining a point on this  
circle, with $\alpha=0$ a matter of choice. Let
$\phi$ be the angle between $\vec{v}^{\; \perp}$ and the transverse  
direction defined by $\alpha$.
The {\it microlensing rate}, $\Gamma$, is the flux of lenses through  
the microlensing tube: 
\begin{equation}
d\,\Gamma =  u_{th} \,
\cos \phi\, v^{\,\perp}\, f(\vec{v})\, d{\vec{v}}\, d\alpha \;
 R_{_E} (x)\, {d n(x)\over dm}\,        dx\, dm \; N(L)\, dL^3 \, ,
\label{dg}
\end{equation}
where $N(L)$ is the number-density of sources and $f(\vec{v})$ is the  
distribution of relative velocities between the lenses and the tube.  
Given the multitude of components of $\vec{v}$, the description of  
$f(\vec{v})$ is laborious.

We work in a non-rotating, centre-of-mass system of the Galaxy. At  
the point $x$ along the line of sight, the microlensing tube ``$\mu  
l t$'', cantilevered from both ends, is moving at a speed:
\begin{equation}
\vec{v}_{\mu l t} = (x/L) \,\vec{v}_s+ (1-x/L)\, \vec{v}_{\odot}\, ,
\label{tube}
\end{equation}
with $\vec{v}_\odot$ the solar system velocity and $\vec{v}_s$ the  
source velocity.

\vskip .3cm
{\it a) Microlensing of LMC and SMC stars by galactic objects}

Consider first the case of observations in the direction of the
Magellanic Clouds, for which it is a very good approximation to
simplify $d\Gamma$ in Eq. (\ref{dg}) in two ways. First, the spread in
distances to the sources is small enough to forsake the $N(L)\, dL^3$
factor. Second, the rotation and dispersion of the source stars 
are negligible. It is therefore sufficient to keep the overall
motion of the Clouds\footnote{ Griest (1991) shows that this motion
corrects the microlensing rates only at the $10 \%$ level.} in the
source velocity $\vec{v}_s$ appearing in Eq. (\ref{tube}), and to
proceed to the discussion of the velocity distribution of the lensing
objects. 

In general, the velocity of the lenses $\vec{v}_{\, lens}$ consists
of a global rotation $\vec{v}_{lrot}$, plus a dispersive component
$\vec{v}_{\, ldis}$ whose distribution we assume to be isotropic and
Maxwellian: 
\begin{equation}
f(v_{\, ldis})= {1 \over (2\pi)^{3/2} \, \sigma_{\, ldis}^3}\; \exp  
\left[{-v_{\, ldis}^2\over 2\, \sigma_{\, ldis}^2}\right]\, ,
\end{equation}
with $\sigma_{\, ldis}$ the dispersion. 
This describes the motion of sources in a Thin or a Thick Disk, while  
for the Halo or Spheroid $\vec{v}_{\, lrot}$ can be neglected.

Finally, the relative velocity $\vec{v}$ between the lens and the  
microlensing tube, occurring in the microlensing rate of  
Eq. (\ref{dg}), is: $\vec{v}=\vec{v}_{\, lens}-\vec{v}_{\mu l  
t}=\vec{v}_{\, lrot} +\vec{v}_{\, ldis}- \vec{v}_{\mu l t}$. Solve  
this expression for $\vec{v}_{\, ldis}$, substitute in Eq. (\ref{dg}),  
change the $v^{\, \perp}$ variable first to $T=R_{_E} /v^{\, \perp}$,  
and then to
$z\equiv R_{_E} /({\sqrt 2}\, \sigma_{lDis} T)$, to obtain:
\begin{equation}
{d\Gamma\over dT}=
u_{th}\, \sigma_{ldis}^2 \,
\int_0^\infty dm \int_0^L dx \; {dn(x)\over dm}\;  z^4\,  
e^{-(z^2+\eta^2)} \; I_0\, (2\eta z)\, ,
\label{dg2}
\end{equation}
where  $\eta\,(x) \equiv\vert \vec{v}_{\mu l t}^{\, \perp}(x) -
\vec{v}_{lrot}^{\, \perp}\vert/({\sqrt 2}\, \sigma_{ldis})$ and $I_0$
is the zeroth-order Bessel function. 

For the sake of elegance, we consider first the idealized case in  
which the observational efficiency as a function of event duration,  
$\epsilon\, (T)$, and the threshold amplification,  $u_{th}(T)$,  
have approximately constant values, $\bar{\epsilon}$ and  
$\bar{u}_{th}$. When discussing  
observations in the next Section, we illustrate the effects of  
lifting the first (and more unrealistic) of these hypothesis.
The second we also revoke
in the discussion of actual data, in Section 6d.

Upon use of the factorization hypothesis, Eq. (\ref{FH}), the total  
rate $\Gamma$ deconvolutes into the product of two terms, one  
depending only on the mass function, the other on the dark density  
profile:
\begin{equation}
\Gamma = \frac{\bar{\epsilon}\,\bar{u}_{th}\, \sigma_{ldis}}  
{\rho_0}\,\sqrt{\frac{8\pi G}{c^2L}}
\int_0^\infty dm\, \sqrt{m}\; \frac{dn_0}{dm} \int_0^L dx  
\,\rho(x)\,\sqrt{x(L-x)}\; \frac{e^{-\eta^2/2}}{\eta}\, M_{_{-1,0}}\,  
(\eta^2),
\label{gamma}
\end{equation}
where $M_{\mu,\nu}$ are the Whittaker functions. The average  
duration, $\langle T \rangle$, and other duration moments, $\langle  
T^n \rangle$, of the $d\Gamma/dT$ distribution also factorize:
\begin{equation}
\langle T^n \rangle \equiv {1\over \Gamma} \int_0^\infty dT\, T^n  
\;{d\Gamma\over dT}
=\left(\frac{2G}{c^2L}\right)^{n\over 2}\frac{1}{\sigma_{ldis}^n}\;  
\frac{{\cal{M}}_n}{ {\cal{M}}_0} \; \frac{{\cal{X}}_n}{{\cal{X}}_0}\,  
,
\label{moments}
\end{equation}
where:
\begin{eqnarray}
{\cal{M}}_n & \equiv& \int_0^{\infty}dm\;  
m^\frac{1+n}{2}\,\frac{dn_0}{dm}\; ,\\
{\cal{X}}_n & \equiv & \Gamma\left(\frac{3-n}{2}\right)
\int_0^L dx\, \rho\, (x)\;\left[x\, (L-x)\right]^\frac{1+n}{2}\,
\frac{e^{-\eta^2/2}}{\eta}\; M_{\frac{n}{2}-1,0}\, (\eta^2)\; .
\label{mom}
\end{eqnarray}
Equations (\ref{gamma} --\ref{mom}) imply that, 
for a known density and velocity distribution of the lensing 
objects, it is possible to extract their mass function --in 
the sense of its moments-- from the observed distribution of event durations
(De R\'ujula, Jetzer \& Mass\'o 1991).

We shall concentrate on the information contained in the total rate  
$\Gamma$ and its $n=1,2$ time-moments, for which we beg to recall  
that $M_{-\frac{1}{2},\, 0} \, (y) = \sqrt{y} \, e^{y/2}$ and   
$M_{0,0}(y)=\sqrt{y}\, I_0(y/2)$ with $I_0$, once again, a Bessel  
function. It will prove useful to combine the $n=1$ and $n=2$  
time-moments into a time dispersion:
\begin{equation}
{\Delta T\over T} \equiv {\sqrt{\langle T^2\rangle -\langle  
T\rangle^2}\over\langle T\rangle}
\, ,
\label{dispersion}
\end{equation}
characterizing the spread of event durations around their mean.

\vskip .3cm
{\it b) Microlensing of LMC stars by dark objects in the LMC}

In the appraisal of LMC ``backgrounds'' to a putative galactic-Halo  
signal, several simplifications are good approximations. The  
microlensing rates are only slightly underestimated by placing the  
sources in the plane of a disk and neglecting its inclination, so as  
to view it from ``above''. The velocity dispersion of the source  
stars can be neglected relative to that of the lenses. Finally, the  
rotation of the sources can also be ignored. It has no effect in the  
case of a dark disk component, for which sources and lenses rotate  
abreast. Moreover, MACHO observes the inner 1.5 kpc of the LMC, while  
the EROS plates portray the inner 2.5~kpc. There, rotation is small;  
its inclusion would only slightly increase the rate and reduce the  
durations of the predicted LMC dark Halo and Spheroid backgrounds.

The microlensing rate and time moments depend on the angle between  
the monitored stars and the centre of the LMC disk, an effect that is  
not negligible for lenses in the LMC itself. Accordingly, we weigh  
the predictions for each observable ${\cal{O}}$ with the radial  
distribution of sources in the LMC disk, $n(R)$, and average over the  
observed fields:
\begin{equation}
\langle{\cal{O}}\rangle = \frac{\int_0^{R_{max}}   
n(R)\;{\cal{O}}(R)\; dR^2} {\int_0^{R_{max}} n(R)\; dR^2}\; .
\label{aver}
\end{equation}
Here, we use $n(R)=e^{-R/h}$, $h=1.6$ kpc, and $R_{max} = h$, roughly  
the field of view of the MACHO collaboration\footnote{ Averaging 
as in Eq. (\ref{aver}) mitigates the uncertainty arising from the 
LMC Halo or Spheroid core radii.}.

\vskip .3cm
{\it c) Bulge observations}

The microlensing predictions for observations in the general  
direction of the galactic centre 
(Paczy\'nski 1991, Griest {{\it et al.}} 
1991) are sensitive to two effects that  
are negligible for the Magellanic Clouds: the dispersion of source  
velocities and their spread of distances along the line of sight. We  
include these effects in our predictions, even though they are  
presumably smaller corrections than those to be made hand in hand  
with the foreseeable improvements of our knowledge of the star  
distributions in, around and towards the Bulge.

Bulge stars have a Maxwellian distribution of velocities $v_{\!  
Bdisp}$ with dispersion $\sigma_{\! B} \simeq 100$~km/s. They probably 
rotate  
jointly at a speed $v_{Brot} \simeq 100$~km/s, but their density  
along the line of sight peaks close to the galactic centre, where  
$v_{Brot}^\perp\sim 0$, so that the effects of rotation on  
microlensing observables can be neglected. The velocity dispersion  
is taken into account by computing the microlensing rate and  
time-moments from the convolution of the $v_{\! Bdisp}$ distribution  
and a function $d\Gamma_{\! B}/dT$ of the same form as that of  
Eq. (\ref{dg2}):
\begin{equation}
\frac{d\Gamma_{\! B}}{dT} = \frac{1}{2\pi\sigma_{\! B}^2}  
\int_0^{2\pi} d\theta
\int_0^\infty dv_{B}\, v_{B} \; \frac{\widetilde{d\Gamma}}{dT} \;  
\exp\left(-\frac{v_B^2}
{2\sigma_{\! B}^2}\right)\; ,
\label{BConv}
\end{equation}
where $v_B = |\vec{v}_{Bdisp}^{\perp}|$
and $\theta$ is the angle between 
$\vec{v}_{Bdisp}^{\perp}$ and $\vec{V} = (1-x/L) \,  
\vec{v}_{\odot}^\perp -\vec{v}_{lrot}^\perp$. The tilde over  
$\widetilde{d\Gamma}/dT$ is to remind one to use Eq. (\ref{dg2}) with  
$\eta\, (x)= |\vec{V}+ (x/L)\,\vec{v}_{\,  
Bdisp}^{\perp}|/(\sqrt{2}\,\sigma_{ldis})$.

The stars monitored in the Bulge are sufficiently disseminated along  
the line of sight to study lifting the approximation wherein they all  
reside at $L=R_\otimes $.
To do this, make yet another convolution of the time distribution of  
Eq. (\ref{BConv}):
\begin{equation}
{d\bar{\Gamma}_{\! B}\over dT} ={\int \left( {d\Gamma_{\! B}}/{
dT}\right)\;N(L)\; L^{-2\,\beta}\; dL^3 \over \int N(L)\;
L^{-2\,\beta}\; dL^3}\; , 
\end{equation}
with $N(L)$ the source number density along the line of sight and  
$L^{-2\,\beta}$ a correction describing the observability of stars of  
given luminosity as a function of distance. We choose $\beta=1$, the  
optimistic endpoint of the range considered by Kiraga and Paczy\'nski  
(1994) and an $N(L)$  distribution for the Bulge with the spheroidal  
shape of Eq. (\ref{Sphe}).



The average over source locations introduces significant corrections  
(a $\sim 25$\% rate increase of and a $\sim 20$\% elongation of event  
durations) only for lenses in the Spheroid. For this distribution the  
most probable lens location is close to the galactic centre,  
enhancing the contribution and Einstein radii of sources at  
$L>R_\otimes$.

\section{Results of Gedanken Observations}

The multiple microlensing rates as functions of event duration  
discussed in Section 4 contain the information needed  
to interpret observations in a hypothetical scenario wherein   
data are plentiful, experimental efficiencies and visibility  
thresholds are uniform and well understood, and the modelling of the  
Galaxy is sufficiently accurate. Before we backtrack a bit from this  
fleckless idealization, it is useful to plot and compare the different  
predictions for $d\Gamma /dT$. To bracket a large range of possibilities,  
we choose the exemplars of lensing bodies with  fixed masses,  
$m_0=0.007M_\odot$ and $m_0=0.1 M_\odot$, the extreme values we  
have adopted.

In Fig. 2, which refers to LMC observations, we portray the  
predicted $d\Gamma/dT$  for  lenses in the galactic Halo, Spheroid,  
and Thin or Thick Disks, as well as for the possible dark populations  
in the LMC Halo, Disk and Spheroid. Figs. 2a and 2b correspond to  
the two chosen extreme values of the lensing mass. The time durations  
of the disk predictions are longer and more peaked than the rest, a  
consequence of global rotation. The expectation for faint disk stars  
is also shown in the figures; in its estimate we used the measured  
mass function (Scalo, 1986), in a mass range extending from the  
H-burning threshold to 0.5 $M_\odot$.

In the ideal case we are discussing, it is easy to generalize the
results of Fig. 2. To predict $d\Gamma/dT$ for mass functions peaked
at some other mass, notice that the height and position of its peaks
respectively scale as $1/m_0$ and $\sqrt{m_0}$, while the shape of the
curves, in a log-log plot, is $m_0$-independent. Wider mass functions
widen the predicted time distributions in ways to be characterized in
detail in the next Section.

Figure 3 is the Baade-window counterpart of Fig. 2. For the models we
consider, the predictions for the other fields observed by OGLE in the
direction of the Bulge are very similar to the ones of Fig. 3. Notice
that, contrary to observation, most of the expected events have time
durations shorter than $\sim$ 15 days, irrespective of the dark
population responsible for them.

\renewcommand{\arraystretch}{1.2}
\begin{table}[t]
\begin{center}
\caption{
Event rate $\Gamma$ (for $10^7$ star-years of observation and $u_{th} =1$) 
and mean event duration $\langle
T\rangle$ (in days) for 100\% efficient microlensing searches in the LMC and
Baade's window. The average distances to the lenses (in kpc) is
$\langle x\rangle$. The symbol  $m_{.05}$ stands for $m/0.05M_\odot$.} 
\vspace{2em}
\begin{tabular}{|c|c|c|c|c|c|}
\hline
Source&Observable& Halo & Spheroid & Thick Disk & Thin Disk\\
\hline
LMC &$\Gamma$ & $92/\sqrt{m_{.05}}$ & $6.8/\sqrt{m_{.05}}$     
& $5.5/\sqrt{m_{.05}}$  & $1.1/\sqrt{m_{.05}}$ \\
 & $\langle T\rangle$  & $15\sqrt{m_{.05}}$  &  
$16\sqrt{m_{.05}}$ & $20\sqrt{m_{.05}}$ & $23\sqrt{m_{.05}}$ \\
 & $\langle x\rangle$  & 15 & 9.2& 3.6 & 1.1\\
\hline
Bulge &$\Gamma$ & $11/\sqrt{m_{.05}}$ & $31/\sqrt{m_{.05}}$ &  
$8.5/\sqrt{m_{.05}}$ & $10/\sqrt{m_{.05}}$ \\
 & $\langle T\rangle$ & $5.8\sqrt{m_{.05}}$ &  
$5.6\sqrt{m_{.05}}$& $11\sqrt{m_{.05}}$ & $11\sqrt{m_{.05}}$ \\
 & $\langle x\rangle$  & 5.6 & 7.7 & 6.2 & 5.7 \\
\hline
\end{tabular}
\end{center}
\end{table}
\renewcommand{\arraystretch}{1}

In Table 1 we list the values of $\Gamma$ and $\langle T\rangle$ for
mass functions peaked at a single mass, for observations in the LMC and in
Baade's Window. We also give the quantities 
\begin{equation} \langle
x\rangle\equiv \frac{1}{\Gamma}\int_0^Ldx {d\Gamma\over dx}x,
\end{equation}
defining, for each dark-mass population, the average distance from the 
solar system
at which microlensing takes place.

\section{Towards the Analysis of Realistic Data}

Microlensing observations are unlikely to accumulate in the near  
future at a rate faster than a few dozen events per year. May such a  
relatively modest body of data answer the burning question of the  
nature of the dark mass of our galactic Halo? We cannot answer this  
question a priori, but we examine simple strategies to eke out the  
information buried in the data. The nine events already gathered by  
OGLE in the Bulge direction are sufficient to start testing these
strategies. 

The first observables of interest are the rate of events $\Gamma$ and
their mean duration $\langle T \rangle$. Their product, or the optical
depth $\tau =(\pi / 2)\, u_{th}\, \Gamma \,\langle T\rangle$, has the
advantage of being, unlike its factors,
independent of the mass function of the lenses (since it is difficult
to assess meaningful error ranges to the various uncertainties, it is
important to individuate the inputs to which the distinct observables
are {\it not} sensitive).

As the data accumulate, more detail of the $T$-distribution will  
become statistically significant, starting with the spread in event  
durations $\Delta T/T$, defined in  Eq. (\ref{dispersion}). For a fixed  
dark-mass distribution, the time moments, unlike $\Gamma$, are  
insensitive to the overall dark-mass normalization, their uncertainty  
stems from that of the assumed
rotational and dispersive velocities. For LMC and SMC observations,  
the predictions for $\Delta T/T$ are quite insensitive to dispersion,  
and are therefore particularly solid for lenses in a spherical Halo  
or Spheroid, that have no collective rotation.

\vskip .3cm
{\it a) LMC observations}

We present predictions for the Halo signal and the various  
backgrounds in the form of two plots, $\Gamma$ versus $\langle  
T\rangle$, and $\Delta T/T$ versus $\langle T\rangle$. The first of
these, for the LMC, is Fig.~4. In Fig.~4a the observational
efficiency is assumed to be ideal, $\epsilon\, (T)=1$, and the
amplification threshold is chosen to be a uniform $u_{th}=1$. Since
the optical depth ($\tau\!\propto\!\Gamma\,\langle T\rangle$) is
independent of the mass function, the predictions for the various
components (in a log-log plot) are straight lines with slope $-1$. The
finite range in $\langle T\rangle$ derives from the assumed bracketed
range of lensing masses\footnote{ The integral of a power-law dark-mass
function is dominated by its endpoints, $m_1$ or $m_2$, for $\alpha>2$
or $\alpha<2$. The corresponding ``critical'' exponent for the rate
$\Gamma$ is a sesquialteral $\alpha=3/2$.}. Faint disk stars would
give the largest values of $\langle T\rangle$, but their rate
contribution to LMC microlensing is negligible. 

In interpreting Fig.~4a and others to come, it is important to  
bethink that
the uncertainties in the various $\Gamma$'s are proportional to those  
in the assumed dark mass densities, and to recall that for the Halo  
signal we have taken a central-value density, while for the various
backgrounds we have made upper-limit estimates. 

In Fig.~4b we illustrate the effects of an  $\epsilon\, (T)\! <\! 1$,
with use of the  efficiency attained by the EROS team during the first
three years of Schmidt plate observations (Bareyre, private
communication). The tilt in the curves stems from that of $\epsilon\,
(T)$, which rises with $T$. The figure shows how the microlensing rate
for a Brown-Dwarf-dominated Halo towers over all backgrounds. 
EROS and MACHO have each accumulated some
$10^7$ star$\cdot$yrs 
of observation, and observed two and three microlensing 
events, respectively. If their efficiencies are comparable and their 
detection thresholds correspond to $u_{th}\sim 1$, the Halo model 
would appear to overestimate the event
rate by half an order of magnitude, within the currently meager statistics.
The
actual results could perhaps be explained by a dominantly 
 dark Spheroid (Giudice, Mollerach \& Roulet 1994) or Thick Disk
(Gould, Miralda-Escud\'e \& Bahcall 1994).

Also shown in Fig.~4 are the predictions for micro--lensing by dark
objects in the LMC itself. The contribution of a Halo component 
would be as much as $\sim 20$ \% of that of the Milky Way Halo.
May the LMC Halo be made of dark massive objects while
that of our Milky Way is dominated by something else? If not, the two
Brown-Dwarf Halo curves must be added accordingly. In the absence 
Brown-Dwarf halos, the contribution of an LMC dark Spheroid may 
not be negligible.

In Fig.~5 we show the results for $\Delta T/T$ versus $\langle
T\rangle$ for the various galactic dark-mass distributions. Fig.~5a
is for a time-independent efficiency and Fig.~5b for the EROS
efficiency cited before. For each dark-mass component, the
indetermination of the mass function $dn_0/dm$ results in predictions
spread over an area of the ($\Delta T/T$, $\langle T\rangle$) plane:
the ``Napoleon hats'' of the figure. For the single-mass distributions
of Eq. (\ref{delta}), $\Delta T/T$ is independent of $m_0$, the lower
boundaries of the hats trace the values of $\langle T\rangle$ as $m_0$
runs from $m_{min}$ to $m_{max}$. The power-law mass functions of
Eq. (\ref{power}) lead to larger values of $\Delta T/T$. The upper hat
boundaries are traced for $m_1=m_{min}$, $m_2=m_{max}$, as $\alpha$
runs from $+\infty$ to $-\infty$, with the top of the hat at $\alpha
=2$. Shorter mass intervals correspond to smaller hats, with the
ensemble of mass functions we consider filling the overall allowed
domain\footnote{ The hypothetical LMC backgrounds are not included in
Fig.~5. Unlike for the rates of Fig.~4, they are not simply additive
and a multitude of combinations is possible, all of which result in 
``hats'' that, relative to the ones of Fig.~5, are somewhat stretched 
towards larger values of $\langle 
T\rangle$ and $\Delta T/T$.}.

We learn from Fig.~5 that, even with relatively modest statistics  
and a realistic efficiency, a measurement of the duration spread of  
the events can help distinguish a Halo signal from the Disk  
backgrounds, whose non-negligible rotation leads to smaller values of  
the predicted $\Delta T/T$. A galactic Halo signal and the Spheroid  
background would have to be disentangled on the basis of their  
different predicted rates, as in Fig.~4.

\vskip .3cm
{\it b) Comparison of SMC and LMC observations}

The MACHO collaboration has started to look also for microlensing
of stars in the SMC. 
The comparison of the microlensing event rates towards the SMC and the 
LMC will provide a useful tool to discriminate among the different models 
of galactic Brown-Dwarf populations. 
We obtain $R_{MC}\equiv \Gamma_{SMC}/\Gamma_{LMC}=1.4$, 1.7, 0.9 and 0.6 
for the Halo, Spheroid, Thick Disk and Thin Disk 
respectively\footnote{ Sackett and Gould (1993) discuss the 
usefulness of $R$ to distinguish
halos of different ellipticity and Gould, Miralda-Escud\'e \& Bahcall
 (1994) to
distinguish between the Disk and and Halo models.}. This
test is particularly good to distinguish between the Spheroid and 
Disks models. The SMC is at a
smaller angle with respect to the galactic centre than the LMC, 
making $R_{MC}$ large for the Halo and Spheroid. The Disk rates 
are very sensitive to latitude, which is slightly larger for 
the SMC than for the LMC, explaining the lower values predicted 
for $R_{MC}$.

The expectations for $d\Gamma /(\Gamma\, dT)$ are similar for the 
LMC and the SMC; the ratio $R_{MC}$ is quite insensitive to the 
assumed mass functions, and almost independent of the observational 
efficiencies,
provided they are similar for the two Clouds. The value of $R_{MC}$ 
is also close to the ratio of optical depths, the quantity 
discussed by Sackett
and Gould (1993) and Gould, Miralda-Escud\'e \& Bahcall (1994). 
Our Thick Disk result 
$R\sim 
0.9$ differs from that of Gould, Miralda-Escud\'e \& Bahcall (1994), 
who did not 
include a radial dependence of the
Disk column density. This is relevant for the Thick Disk, for which 
microlensing occurs at $\langle x \rangle = 3.6$ kpc, 
comparable to the Disk's scale length. 

To obtain $R_{MC}$ we neglected possible
contributions from dark lensing objects around the Magellanic Clouds
themselves. If such contributions were sizable, estimates of
$R_{MC}$ would actually require a detailed
knowledge of the lens distribution in those galaxies,
not a readily available piece of information.

\vskip .3cm 
{\it c) Bulge observations} 

The predictions for microlensing rates as functions of average times,
for observations in the Baade window, are given in Fig.~6. Fig.~6a
is for a perfect efficiency, while in Fig.~6b we have used $\epsilon(T)$ as
 quoted by the OGLE team (Udalski {\it et al.} 1994). Dim objects in the
Spheroid and Disk populations result in larger event rates than those
predicted for the galactic Halo, the transpose of the situation for
the LMC depicted in Fig.~4. 


In Fig.~7 we show the $\Delta T/T$ versus $\langle T\rangle$ 
plane for the Bulge. Unlike for the LMC case of Fig.~5, it seems much
more difficult to discriminate among the different galactic components
with measurements of $\langle T\rangle$ and $\Delta T/T$. Also, the
effect of the experimental OGLE efficiency strongly modifies the
picture, as shown in Fig.~7b. This is caused by the rapidly falling
efficiency at small time durations, where the theoretical time
distributions are peaked. 

The known faint Disk stars complicate the analysis of microlensing 
in the Bulge direction. If the other dark populations have mass 
functions not peaked close to
0.1 $M_\odot$, two distinct peaks may arise in the distribution 
of event durations, otherwise the contribution of faint Disk stars 
must be handled theoretically.

\vskip .3cm
{\it d) An inspection of the OGLE data}

In $1.7\times 10^6$ star-yrs of observation, the OGLE group has seen 9
events, one of which may be  a long-period variable star.
Figure (6) shows that the remaining 8 events are in excess of 
the individual predictions of 
any of the galactic components, but may be compatible with the 
largest expectation of $\sim
5$ events, obtained by adding 
the rates for the Spheroid, either of the dark Disk models, 
and the faint stars. The observed $\langle T\rangle\simeq 20$~days 
is in the upper end of 
the expectations, that do not exceed that value for any of the 
dark-mass models. 
How significant are these uncomfortable indications? Poissonian 
statistics suffice to compare
the numbers of expected and observed events, but a probabilistic test
of a time distribution necessitates a hypothesis on its shape, plus
some extra labour. 

One of the OGLE events is best interpreted as due to microlensing 
by a binary
object; it is difficult to include it in an analysis of time durations.
The event compatible with a variable star is suspicious and we also
discard it, to be left with 7 events. Seven is too small a number for 
a conventional $\chi^2$ test of a binned distribution to be safe, we 
must resort to a
Kolmogorov--Smirnov test of the cumulative probability of the observed
distribution to agree with the expectation\footnote{ In the cumulative 
probability we use steps at each measured event duration $T^i$ 
proportional to the inverse of its individual threshold $u_{th}^i$, 
reported in Udalski {\it et al.} (1994).}. 
Since the observed
durations are large, the expectations least likely to ``fail'' are
those corresponding to mass functions peaked at a single large mass, 
justifying the use of a single
mass-ansatz. The background of faint stars also corresponds to large
durations and we add it to each of the individual dark-mass models. 

The results of the above exercise are shown in Fig.~8, where 
the Kolmogorov-Smirnov probability as a function of lensing 
mass is shown for each dark population model. The Spheroid and, 
to a lesser extent, the Halo model fare badly for masses below 
the hydrogen-burning limit. The Disk models survive this test 
with flying colours, 
but fail the rate test of Fig.~6b: in either of them one would 
expect less than two events for the currently accumulated 
statistics, the probability of observing 8 or 9 events is totally 
negligible. Thus the unavoidable need to complicate former 
models of the inner galaxy.

To face these problems, the MACHO observers (Alcock {\it et al.} 1994)
suggest a very centrally-singular Halo, a maximal Disk and the
possible contribution of source stars lying behind the Bulge, with the
greater distance to the lensing object increasing the lensing
probability and time scale. Members of the OGLE collaboration
(Paczy\'nski {\it et al.} 1994) advocate the effects of the {\it
galactic Bar}, a cigar-like ensemble of stars whose elongated axis
would point towards us from the neighbourhood of the galactic centre,
at an angle of only $\sim 15^\circ$ relative to the line of sight
(Binney {\it et al.} 1991). Lensing by dim stars in the Bar would
contribute to the large observed event rate; their small apparent
transverse velocities would result in relatively long-duration events.

\bigskip
\section{Conclusions}

Microlensing observations are the best current tool to search for 
compact lumps of
baryonic dark-matter in the Galaxy. The observational campaigns 
were designed to ascertain the extent to which the galactic Halo 
consists of massive astrophysical objects or, by exclusion, of a 
more elusive substance. Infrared searches, that we have not 
discussed, are another tool to locate nearby individual Brown 
Dwarfs, or their collective glow in another galaxy.

It is becoming increasingly clear that the microlensing observations 
are sensitive to ``backgrounds'' of dark objects residing in galactic 
components other than the Halo, such as the Spheroid and the Thick or 
Thin Disks. We have presented a detailed description of the microlensing 
Halo signal and the various backgrounds, an analysis designed to 
accomodate the foreseeable scarcity of data. Our aim is to pin down 
the likely location of the lensing objects, as well as to extract the 
first indications of what their mass function may be. 

The observations of the Bulge are intriguing. We have analysed the 
statistical significance of the discrepancy between the OGLE observations 
and the expectations for simple dark-mass models and known faint stars, 
or combinations thereof. In agreement with previous authors 
(Alcock {\it et al.}, 1994;
Paczy\'nski {\it et al.} 1994), we conclude that the earlier understanding 
of the inner galactic realm must be revised. In this connection,
microlensing observations with a good sensitivity for short-duration events 
(10 days or less), for which the expected rates are large, would be very 
useful. So would the comparison of observations in ``windows'' located at 
different latitudes and relative angles to the galactic centre.

Needless to emphasize, even a Brown-Dwarf-dominated galactic Halo would 
contribute very little to microlensing of Bulge stars, so that observations 
in that direction are not decisive to the question of the Halo constituency, 
though they constitute a handle on various microlensing ``backgrounds''.

We have argued that for the microlensing of LMC stars, the comparison 
of data and expectations for the rate, mean event duration and time 
dispersion should suffice to disentangle the galactic component to 
which the Brown-Dwarfs belong. A comparison of LMC and SMC rates may 
also come in handy.

To summarize, microlensing observations of the galactic Bulge
are already significantly contributing to our astrophysical lore; 
very soon observations of the Magellanic Clouds ought to add to 
our knowledge of cosmology. 
The question of the nature of the Halo of our galaxy is still wide 
open, but the prospects for continuing progress appear to be excellent.

\section{References}

\noindent Alcock C. {{\it et al.}}, 1993. Nature {\bf 365}, 621. 

\noindent Alcock C. {{\it et al.}}, 1994. Preprint. 

\noindent Aubourg E. {{\it et al.}}, 1993. Nature {\bf 365}, 623. 

\noindent Bahcall J. N., Schmidt M. \& Soneira R. M., 1983. Astrophys.
J. {\bf 265}, 730. 

\noindent Bahcall J. N., 1984. Astrophys. J. {\bf 276}, 169. 

\noindent Bahcall J. N. {{\it et al.}}, 1994. Preprint.

\noindent Binney, J. \& Tremaine, S., 1987. Galactic Dynamics,
Princeton University Press, Princeton. 

\noindent Binney J. {\it et al.}, 1991. Mon. Not. R. astr. Soc. {\bf 252},
210. 

\noindent Blanco V. M. \& Tendrup D.  M., 1989. Astronom. J. {\bf 98},
843. 

\noindent Blitz L. \& Spergel D. N., 1991a. Astrophys. J. {\bf 370},
205. 

\noindent Blitz L. \& Spergel D. N., 1991b. Astrophys. J. {\bf 379},
631. 

\noindent Caldwell J. A. R. \& Ostriker J. P., 1981. Astrophys. J. {\bf
251}, 61. 

\noindent Carswell R. F. {\it et al.}, 1994. Mon. Not. R. astr. Soc., in
press. 

\noindent Chandrasekhar, S., 1942. Principles of Stellar Dynamics, The
University of Chicago Press, Chicago. 

\noindent De R\'ujula A., Jetzer P., \& Mass\'o E., 1991. Mon. Not. R.
astr. Soc. {\bf 250}, 348. 

\noindent De R\'ujula A., Jetzer P., \& Mass\'o E., 1992. Astron.
Astrophys. {\bf 254}, 99. 

\noindent Einstein A., 1936. Science {\bf 84}, 506. 

\noindent Fich M. \& Tremaine S., 1991. Ann. Rev. Astron. 
Astrophys. {\bf 29}, 409.

\noindent Freeman K. C., 1987. Ann. Rev. Astron. Astrophys. {\bf 25},
607. 

\noindent Frieman, J. \& Scoccimarro, R., 1994. Preprint. 

\noindent Gilmore G. \& Reid N., 1983. Mon. Not. R. astr. Soc. {\bf
202}, 1025. 

\noindent Giudice G.F., Mollerach S. \& Roulet E., 1994. Phys. Rev. 
D in press. 

\noindent Gould A., 1990. Mon. Not. R. astr. Soc. {\bf 244}, 25. 

\noindent Gould, A., 1993. Astrophys. J. {\bf 404}, 451. 

\noindent Gould A., Miralda-Escud\'e J. \& Bahcall J. N., 1994.
Astrophys. J. {\bf 423}, L105. 

\noindent Graboske H. C. \& Grossman A. S., 1971. Astrophys. J. {\bf
170}, 165. 

\noindent Griest K., 1991. Astrophys. J. {\bf 366} 412. 

\noindent Griest K. {{\it et al.}}, 1991. Astrophys. J. {\bf 372}, L79. 

\noindent Hughes S. M. G., Wood P. R. \& Reid N., 1991. {\it IAU
Symposium ``The Magellanic Clouds"}, 15, eds. R. Haynes and D. Milne. 

\noindent Kent S. M., 1992. Astrophys. J. {\bf 387}, 181. 

\noindent Kerins E. J. \& Carr B. J., 1994. Mon. Not. R. astr. Soc. 
{\bf 266}, 775. 

\noindent Kiraga M. \& Paczy\'nski B., 1994. Astrophys. J. Letters, in
press. 

\noindent Kuijken K. \& Gilmore G., 1989a. Mon. Not. R. astr. Soc. {\bf
239}, 571. 

\noindent Kuijken K. \& Gilmore G., 1989b. Mon. Not. R. astr. Soc. {\bf
239}, 605. 

\noindent Low C. \& Lynden-Bell D., 1976. Mon. Not. R. astr. Soc. {\bf
170}, 367. 

\noindent Ostriker J. P. \& Caldwell J. A. R., 1982. In {\it Dynamics
and Structure of the Milky Way}, ed. W. L. H. Shuter, Reidel,
Dordrecht. 

\noindent Paczy\'nski B., 1986. Astrophys. J. {\bf 304}, 1. 

\noindent Paczy\'nski B., 1991. Astrophys. J. {\bf 371}, L63. 

\noindent Paczy\'nski B. {\it et al.}, 1994. Preprint. 

\noindent Palla F., Salpeter E.E. \& Stahler S.W., 1983. Astrophys. J.
{\bf 271}, 632. 

\noindent Rich R. M., 1990. Astrophys. J. {\bf 362}, 604. 

\noindent Richer H. B. \& Fahlman G. G., 1992. Nature {\bf 358}, 383. 

\noindent Rohlfs K. \& Kreitschman J., 1988. Astron. Astrophys. {\bf
201}, 51. 

\noindent Sackett P. \& Gould A., 1993. Astrophys. J. {\bf 419}, 648. 

\noindent Sackett P. {\it et al.}, 1994. Nature {\bf 370}, 441.

\noindent Sahu, K., 1994. Preprint. 

\noindent Scalo J. M., 1986. Fund. of Cosmic Phys. {\bf 11}, 1. 

\noindent Schommer R. A. {{\it et al.}}, 1992. Astron. J. 103, 447. 

\noindent Songaila A. {\it et al.}, 1994. Nature {\bf 368}, 599. 

\noindent Storm J. \& Carney B. W., 1991. {\it IAU Symposium ``The
Magellanic Clouds"}, 15, eds. R. Haynes and D. Milne. 

\noindent Udalski, A. {{\it et al.}}, 1994. Acta Astronomica {\bf 44}, 165.

\noindent Walker T. P. {\it et al.}, 1991. Astrophys. J. {\bf 376}, 51. 

\noindent Westerlund B. E., 1991. {\it IAU Symposium ``The Magellanic
Clouds"}, 15, eds. R. Haynes and D. Milne. 

\noindent Wu X.-P., 1994. Preprint. 

\section{Figure Captions}

\noindent{\bf Figure 1:} 
On a plane orthogonal to the Galaxy's equatorial plane, this figure
shows the mass density contour lines of $10^{-2}M_\odot /\mbox{pc}^3$
and $10^{-3}M_\odot /\mbox{pc}^3$ for the Halo (solid line), the
Spheroid (dashed line), the Thick Disk (solid line), and the Thin Disk
(dotted line). The location of the Sun is indicated by an asterisk.
The Magellanic Clouds lie above the plane, as indicated. 

\noindent{\bf Figure 2:}
The microlensing event distributions in time durations for
observations towards the LMC predicted by the 
various Milky Way and LMC components. All dark objects are
assumed to have the same mass, $M =0.007M_\odot$ in Fig.~2a
and $M =0.1M_\odot$ in Fig.~2b.

\noindent{\bf Figure 3:}
The analog of Fig.~2 for
observations towards the galactic Bulge.

\noindent{\bf Figure 4:}
The microlensing event rate, $\Gamma$, versus
the average event duration, $\langle T\rangle$, for observations
towards the LMC predicted by the various Milky Way and LMC components,
with the mass functions varied as in the text and $u_{th}=1$.
The asterisk shows the prediction of known faint-disk stars.
The efficiency $\epsilon (T)$ is unity in Fig.~4a
and equal to the EROS efficiency for Schmidt plate observations
(Bareyre, private communication) in Fig.~4b.

\noindent{\bf Figure 5:}
The analog of Fig.~4 for the allowed domains in the
($\langle T\rangle$, $\Delta T/T$) plane
of average time duration versus time dispersion.  

\noindent{\bf Figure 6:}
The microlensing rate, $\Gamma$, versus
the average event duration, $\langle T\rangle$, for observations
towards the galactic Bulge 
predicted by the different Milky Way components,
with the  mass functions varied as in the text and $u_{th}=1$. The
asterisk shows the prediction of known faint disk stars. The
experimetal efficiency $\epsilon (T)$ is unity in Fig.~6a and equal to
the OGLE efficiency (Udalski {\it et al.} 1994) in Fig.~6b. 

\noindent{\bf Figure 7:}
The analog of Fig.~6 for the allowed domains in the
($\langle T\rangle$, $\Delta T/T$) plane
of average time duration versus time dispersion.  

\noindent{\bf Figure 8:} Kolmogorov-Smirnov probability that the OGLE 
observed distribution  of event durations be consistent with the 
predictions of the different Milky Way dark components, as function of 
the assumed single lens mass. The faint 
star contribution has been added to each distinct model.

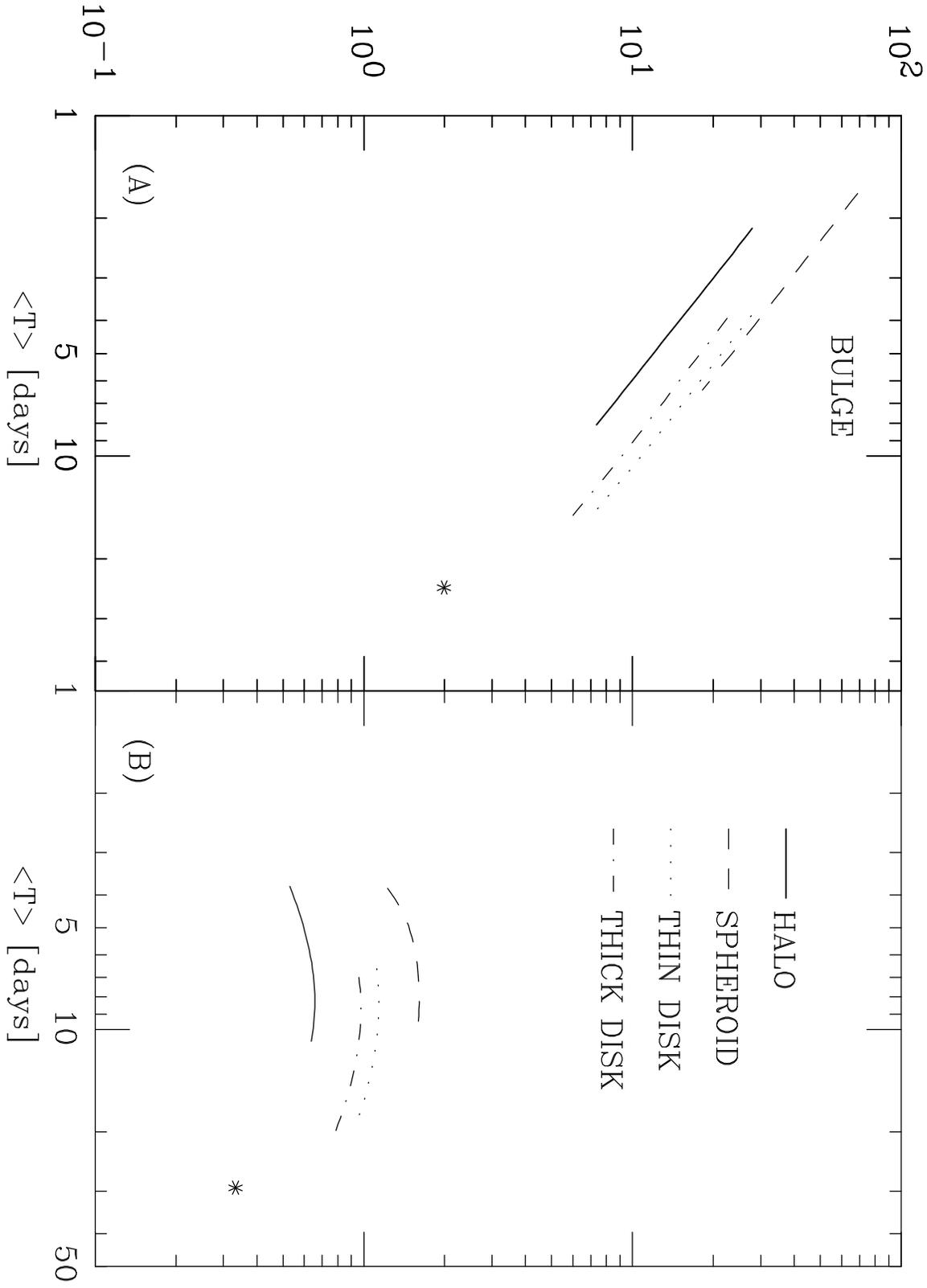
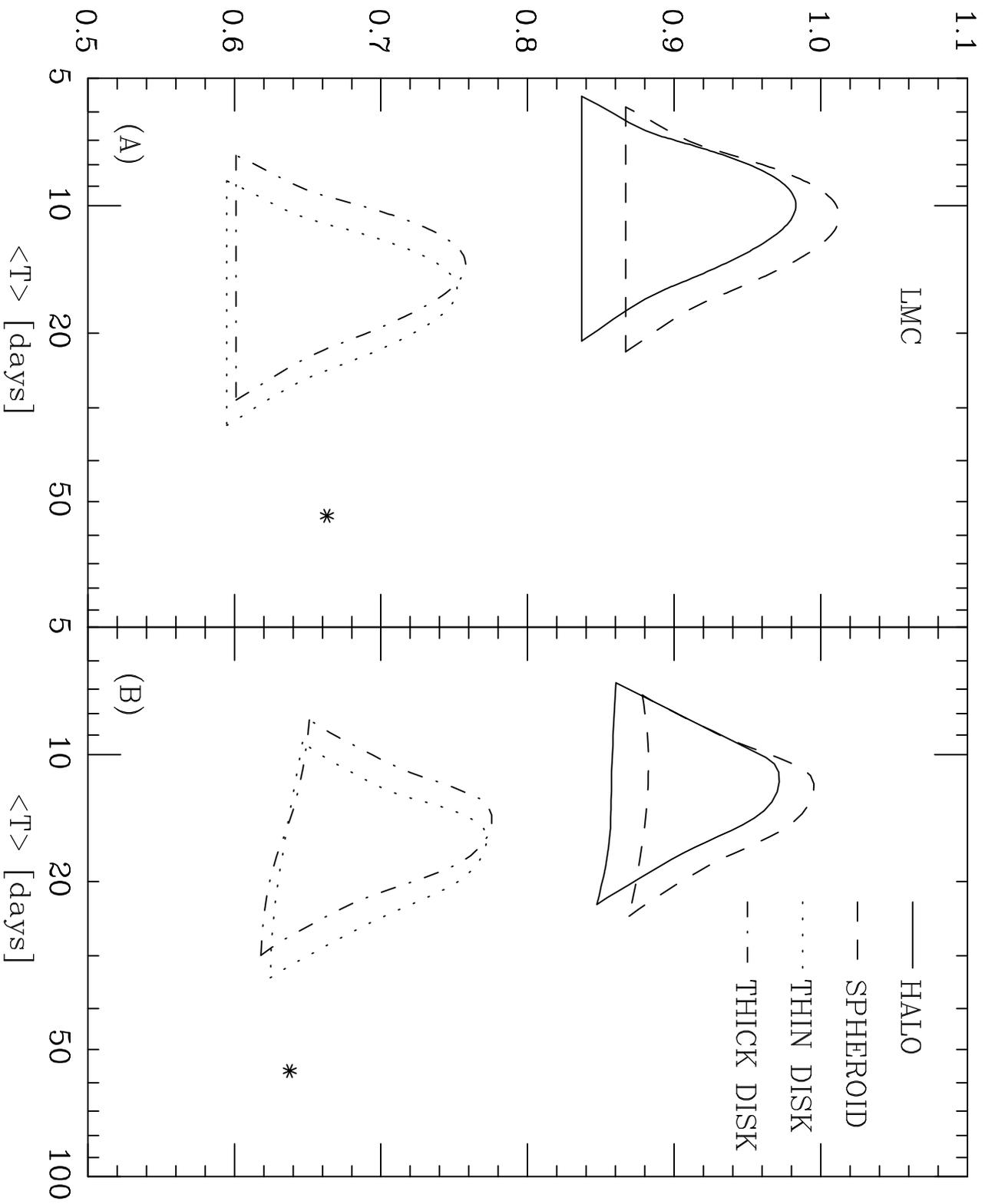
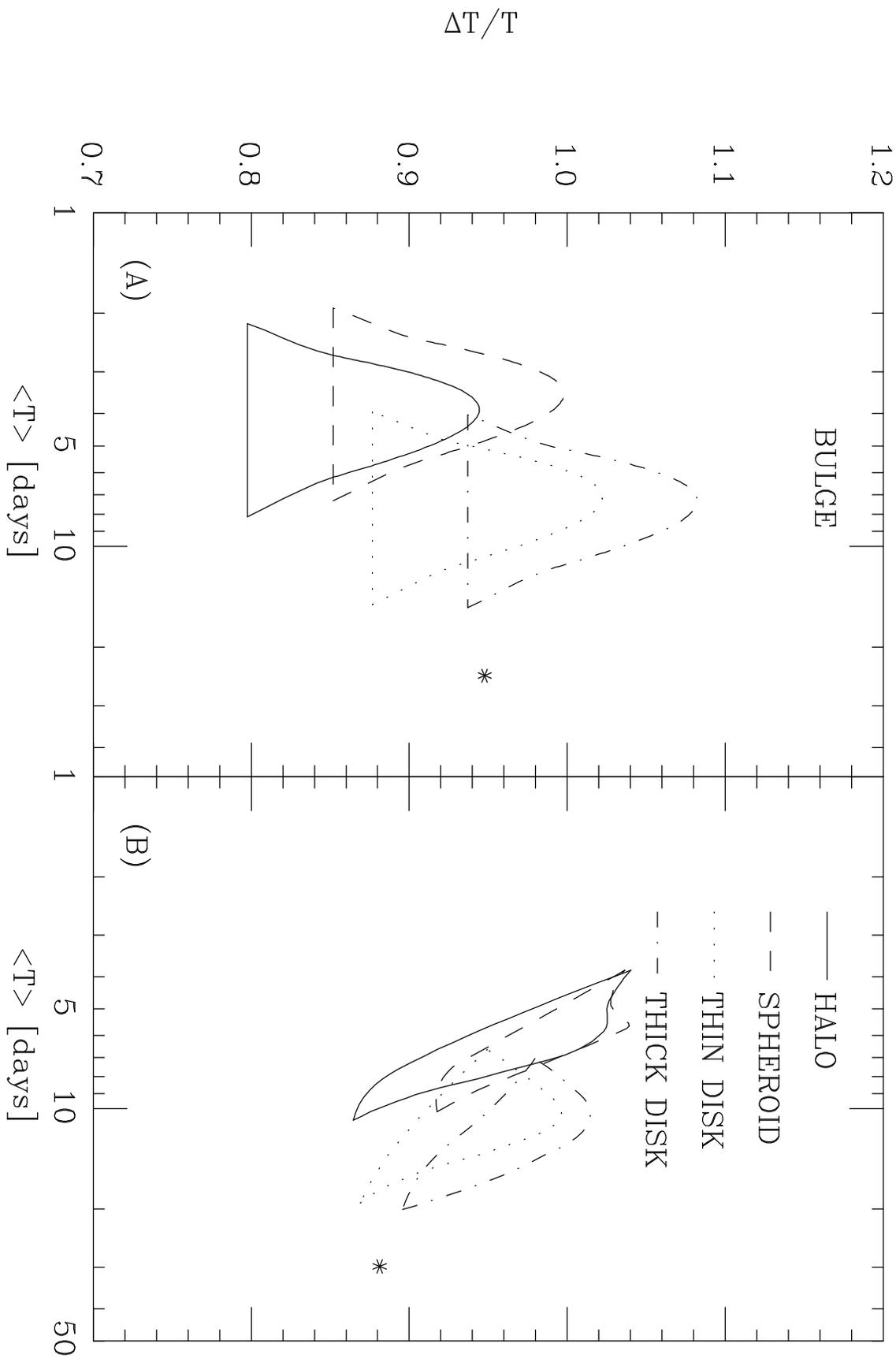
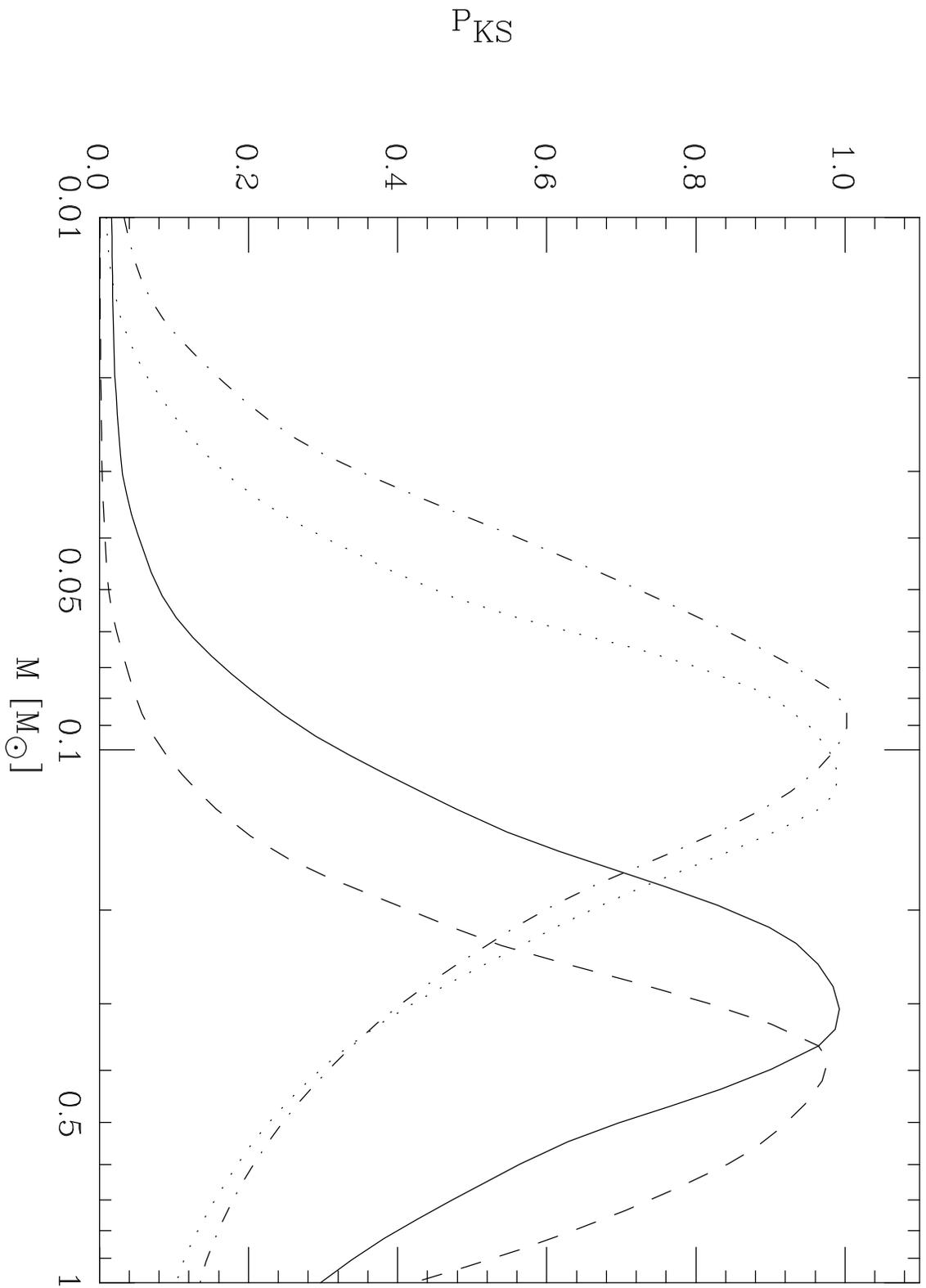
\end{document}